\documentclass{aa}  

\usepackage{graphicx}
\usepackage{txfonts}
\usepackage{epsfig}
\usepackage{longtable}
\usepackage{lscape}
\usepackage{amssymb} 
\usepackage[colorlinks=true,citecolor=blue]{hyperref} 
\usepackage{refcount}

\def \sw {{\em Swift}}

\def \hcm {\hbox {\ifmmode $ atom cm$^{-2}\else atom cm$^{-2}$\fi}}

\def \apj {ApJ}
\def \apjl {ApJL}
\def \apjs {ApJS}\def \aap {A\&A}

\def \mnras {MNRAS}
\def \ssr {SSR}

\begin{document} 

\title{Constraining duty cycles through a Bayesian technique} 
\titlerunning{Uncertainties on duty cycles} 
\authorrunning{Romano et al.}

\author{P.\ Romano\inst{1}
           \and
            C.\ Guidorzi\inst{2}
           \and
            A.\ Segreto\inst{1}
         \and
         L.\ Ducci \inst{3,4} 
         \and
         S.\ Vercellone\inst{1} 
         }
   \institute{INAF, Istituto di Astrofisica Spaziale e Fisica Cosmica - Palermo,
              Via U.\ La Malfa 153, I-90146 Palermo, Italy \\
              \email{romano@ifc.inaf.it}
              \and
              Dipartimento di Fisica e Scienze della Terra, Universit\`a{} di Ferrara, Via Saragat 1, I-44122 Ferrara, Italy 
                \and 
             Institut f\"ur Astronomie und Astrophysik, Eberhard Karls Universit\"at, 
             Sand 1, 72076 T\"ubingen, Germany
             \and 
             ISDC Data Center for Astrophysics, Universit\'e de Gen\`eve, 16 chemin d'\'Ecogia, 1290 Versoix, Switzerland
              }

\date{Received 11 May 2014; accepted 2014 October 15 }

\abstract{
The duty cycle (DC) of astrophysical sources is generally defined as the fraction of time 
during which the sources are active. It is used to both characterize their central engine 
and to plan further observing campaigns to study them. 
However, DCs are generally not  provided with statistical uncertainties, since 
the standard approach is to perform Monte Carlo bootstrap simulations to evaluate them,
which can be quite time consuming for a large sample of sources. 
As an alternative, considerably less time-consuming approach, we
derived the theoretical expectation value for the DC and its error for sources whose state 
is one of two possible, mutually exclusive states, inactive (off) or flaring (on), 
as based on a finite set of independent observational data points. 
Following a Bayesian approach, we derived the analytical expression for the posterior,
the conjugated distribution adopted as prior, and the expectation value and variance. 
We applied our method to the specific case of the inactivity duty cycle 
(IDC) for supergiant fast X--ray transients, 
a subclass of flaring high mass X--ray binaries characterized by large dynamical ranges. 
We also studied IDC as a function of the number of observations in the sample. 
Finally, we compare the results with the theoretical expectations. 
We found excellent agreement with our findings based on the standard bootstrap method.  
Our Bayesian treatment can be applied to all sets of independent observations of 
two-state sources, such as active galactic nuclei, X--ray binaries, etc.  
In addition to being far less time consuming than bootstrap methods, 
the additional strength of this approach becomes obvious when considering a well-populated 
class of sources ($N_{\rm src} \geq 50$) for which the prior can be fully characterized 
by fitting the distribution of the observed DCs for all sources in the class, so that, 
through the prior, one can further constrain the DC of a new source
by exploiting the information acquired on the DC distribution derived from the other sources. 
}   

\keywords{Methods: statistical -- Methods: numerical-- Methods: observational -- X-rays: binaries}

\maketitle

\setcounter{table}{0}  
 \begin{table*}  
 \tabcolsep 4pt         
 \begin{center}         
 \caption{Source sample properties and comparison of measured IDCs with  Bayesian estimates and MC simulations. }
 \label{sfxtsims:tab:final}       
 \begin{tabular}{lrrrr ccc rc} 
 \hline 
 \hline 
\noalign{\smallskip} 
  Source                        & Orbital & \multicolumn{2}{c}{Observation}  &IDC\tablefootmark{a}                                                                                     
                                    &  \multicolumn{3}{c}{\hspace{12pt} Bayesian method}   
                                  & \multicolumn{2}{c}{Monte Carlo simulations}\\  
                              & period  & $N$   & Type        &          & \multicolumn{3}{c}{\hspace{12pt}  confidence intervals \tablefootmark{b}}     
                            &  \hspace{12pt} $\overline{IDC_{\rm sim}}\pm s_{\rm sim}$\tablefootmark{c}  &  \hspace{12pt} $S_{a}$\tablefootmark{d}  \\ 
                              & (d)       &            &        & (\%)  & 68.3\,\%  & 95.4\,\%   &  99.7\,\%     & (\%)   & \\   
 \noalign{\smallskip}
 \hline
 \noalign{\smallskip} 
IGR~J08408$-$4503    & --              & 77   & Y    & 67.2   & 61.5--72.1     & 55.8--76.8     & 50.1--81.2 &     $67.3\pm5.6$  & 40 \\  
IGR~J16328$-$4726    &$10.076$    & 94   & Y  & 61.0     & 55.8--65.8     & 50.7--70.4     & 45.5--74.8 &        $61.0\pm5.6$ & 40 \\  
IGR~J16465$-$4507    &$30.243$    & 61   & Y   &   5.1    & 3.5--9.5         & 1.8--13.9      & 0.8--19.2  &        $5.2\pm2.9$  & 40 \\  
IGR~J16479$-$4514    &$3.3193$    &139  & Y  & 19.4     &16.5--23.2     & 13.6--26.9     & 11.0--30.9 &      $19.4\pm3.6$ & 80 \\  
XTE~J1739$-$302       &$51.47$      & 181 & Y   & 38.8    & 35.3--42.5     & 31.9--46.2     & 28.5--49.9 &     $39.0\pm4.7$   & 70 \\  
IGR~J17544$-$2619   &$4.926$       & 138 & Y  & 54.5    & 50.2--58.6     & 46.0--62.7     & 41.8--66.7 &     $54.5\pm5.3$ & 50\\ 
AX~J1841.0$-$0536    & --              & 87   & Y   & 28.4    & 24.1--33.7     & 19.8--38.9     & 16.0--44.2 &     $28.5\pm5.6$& 40\\   
  \noalign{\smallskip}
IGR~J16418$-$4532    &$3.73886$  & 15   & O  & 11.0    & 7.2--24.1      & 3.0--36.1      & 0.9--49.1 &     $11.3\pm8.0$ & --\\  
IGR~J17354$-$3255    &$8.448$      & 22   & O  & 33.4    & 25.1--44.5     & 17.2--54.9     & 11.0--64.8 &     $33.3\pm10.4$ & --\\  
IGR~J18483$-$0311    &$18.545$    & 23   & O  & 26.6    & 19.5--37.5     & 12.7--47.6     & 7.6--57.8 &    $26.7\pm9.4$ & --\\  
  \noalign{\smallskip}
  \hline
  \end{tabular}
  \end{center}
\tablefoot{
\tablefoottext{a}{From Eq.~\ref{sfxtsims:eq:IDC} \citep[see][and references therein]{Romano2014:sfxts_paperX}.}
\tablefoottext{b}{Theoretical confidence intervals of IDC (Sect.~\ref{sfxtsims:statistics}, Eq.~\ref{sfxtsims:eq:posterior2}). }    
\tablefoottext{c}{Simulated sample mean and standard variance 
(Sect.~\ref{sfxtsims:mcboot_idc}, $M=10^4$ data sets drawn from the observed sample of size $N$).}   
\tablefoottext{d}{Minimum number of observations required for an IDC with the desired accuracy 
(Sect.~\ref{sfxtsims:mcboot_idc_s}, $M=10^4$ data sets, drawn from a sample of size $S=10, 20, 30, ..., N$).}
} 
  \end{table*}

	\section{Introduction\label{sfxtsims:intro}}

In astrophysics it is often crucial to determine the duty cycle  (DC)  of a source, 
or a class of sources, in order to understand both their central engines 
and to plan additional observing campaigns aiming at best studying them. 
Generally, the DC is defined as the fraction of time, usually expressed in percentages, 
during which the source is active, or 
\begin{equation} 
\label{sfxtsims:eq:DC}
{\rm DC}= T_{\rm active} / T_{\rm Tot} \, ,  
\end{equation}
where $T_{\rm active}$ is the time spent above some instrumental threshold or 
some scientifically interesting flux value, and $T_{\rm Tot}$ is the total exposure. 
In the case of periodic sources, such as classical X--ray binaries, $T_{\rm active}$ is generally 
the time during which an $n$-$\sigma$ detection  is achieved ($n$ being 3 or 5, depending on the 
detection method), and $T_{\rm Tot}$ is the orbital period $P_{\rm orb}$ or the 
spin period $P_{\rm spin}$ \citep[e.g.][]{Henry1969,Fragos2009,Knevitt2014}. 

For active galactic nuclei (AGNs), the DC is often defined as the fraction of time a source 
spends in a flaring state, that is, at $n$ times the average flux, $\overline{F}$, with 
$n$ being a small number, depending on the purpose of the study 
\citep[e.g.][]{Jorstad2001:DC,Vercellone2004:DC,Ackermann2011:FermiAGNCat2}. 
For example, in \citet[][]{Vercellone2004:DC} the DC is defined as 
$\chi = \frac{\tau}{\tau + T}\;, $
where  $T$ is the time spent in a low flux level ({\it off state}) and $\tau$ is the time spent in a high
flux level ({\it on state}),  defined by $HSN = \sum_{i=1}^{n} C_{i}$, where 
$C_{i} =  1$ if $F_{i} \ge 1.5 \times \overline{F}$ and $C_{i}  =  0$  otherwise. 

Alternatively, when a source shows a very large dynamical range (a few orders of magnitude), 
more can be inferred about its nature by considering the {\it inactivity duty cycle}  
 \citep[IDC,][]{Romano2009:sfxts_paperV}  defined as the time a source remains undetected 
down to a certain flux limit $F_{\rm lim}$, 
\begin{equation} 
\label{sfxtsims:eq:IDC}
{\rm IDC}= \Delta T_{\Sigma} / [\Delta T_{\rm tot} \, (1-P_{\rm short}) ] \, ,  
\end{equation}
where  $\Delta T_{\Sigma}$ is the sum of the exposures accumulated in all observations 
where only a 3$\sigma$ upper limit was achieved, 
$\Delta T_{\rm tot}$ is the total exposure accumulated, 
and $P_{\rm short}$ is the percentage of time lost to short observations that 
need to be discarded in order to differentiate between non-detections 
due to lack of exposure from non-detections due to a true low flux state. 

Since DCs (and IDCs) are integral quantities depending on the total observing time 
and the total time spent above (or below) a given flux threshold, they 
are implicitly dependent on the instrumental sensitivity,  
observing coverage, and the characteristic source variability timescales. 
The implicit assumption is that, in order to obtain a meaningful DC, the observations used to
calculate them are independent, that is, each observation is not triggered by the 
previous ones. This is the case, for example, of monitoring programmes whose 
monitoring pace and exposures are defined a priori and do not depend on the
source state.

In this paper we determine the theoretical expectation value of DC and its error.
We then consider one specific case, the IDCs measured from ten \sw\ \citep{Gehrels2004} 
X--ray Telescope \citep[XRT,][]{Burrows2005:XRT} observing campaigns on 
supergiant fast X--ray transients (SFXTs),  a subclass of high mass X--ray binaries 
known for their rapid hard X--ray flaring behaviour and 
large dynamical range (up to 5 orders of magnitude), 
and compare the theoretical expectations with both the observed values 
and with those obtained from Monte Carlo simulations. 
We also evaluate how the IDC varies as a function of the number of observations available 
and estimate how many observations are required to obtain an IDC within a desired accuracy. 
Finally, we supply the reader with useful $R$--language \citep[][]{Rlanguage:2014}, IDL, and C--language
procedures to calculate several confidence intervals (c.i.) on the DC estimate for a given source.

\begin{figure*}
\vspace{-10truecm}
\hspace{-0.5truecm}
\centerline{\includegraphics*[width=20cm]{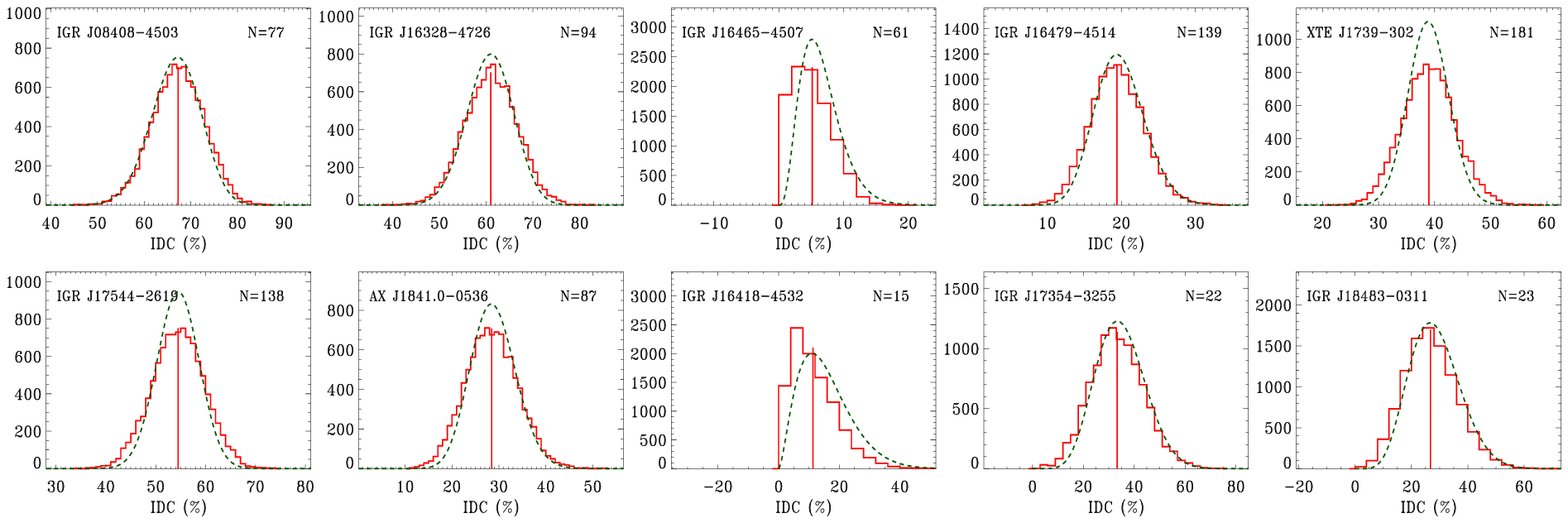}} 
\vspace{-12truecm}
\caption{
Distribution of IDC values derived from $10^4$ bootstrap 
simulations  (red), each drawn from a sample of size $N$. 
The solid vertical line marks the simulated sample mean from Eq.~(\ref{sfxtsims:eq:IDCsim}). 
The dashed (green) lines are the curves described by Eq.~(\ref{sfxtsims:eq:posterior2})
in the case of a uninformative prior ($a=b=1$). 
}
\label{sfxtsims:fig:distro_idc_M10000}
\end{figure*}

\section{Statistical estimate of the duty cycle \label{sfxtsims:statistics}}

We consider one source for which $N$ independent observations were collected 
and for which the DC was calculated as described in Sect.~\ref{sfxtsims:intro}. 
In the following we estimate the DC, that we hereafter define as $\mu$, but the 
formalism is unchanged for the case of the IDC which we consider in Sect.~\ref{sfxtsims:mcboot_idc}.    
In all generality, the stochastic variable {\it state of the source} can be seen as a 
discrete random variable that can take only one of two possible, mutually exclusive states, 
active (off) and flaring  (on), so that $\mu$ is the probability of finding the source 
active in a given casual pointing. 
After $N$ observations, the probability of finding the source active $m$ times is
given by the binomial 
\begin{equation}
\label{sfxtsims:eq:binomial}
{\rm Bin} (m \,| \, N, \mu) = \binom{N}{m} \, \mu^{m} \,  (1-\mu)^{N-m} \, ,
\end{equation}
with an expectation value $E\{m\}=\mu N$ and variance var$\{m\}=N \mu (1 - \mu)$.
Once $N$ and $m$ are known, 
where  $m=N \mu_{\rm est}$ and $\mu_{\rm est} $ is the DC measured from the $N$ observations,  
then the problem becomes estimating the statistical offset of $\mu$ from $\mu_{\rm est}$.
From the central limit theorem $\mu_{\rm est}$ is normally distributed in the
limit of large values of $m$ and $N-m$ with $E\{\mu_{\rm est}\}=\mu$ and
$\sigma\{\mu_{\rm est}\}=\sqrt{\mu \, (1-\mu)/N}$.

Hereafter, we adopt a Bayesian treatment, in which $\mu$ is treated as a random variable
whose probability density function (PDF) depends on the observed values for $N$ and $m$. 
From the Bayes theorem, the posterior distribution is proportional to the product of the
likelihood and the prior function, and 
the posterior distribution is to all intents and purposes a PDF of the random variable $\mu$
given the observed values for $N$ and $m$,   
\begin{equation}
\label{sfxtsims:eq:posterior}
{\rm P} \, (\mu \,| \, N, m)  \propto  {\rm P} (m \, | \, N , \mu) \cdot p \, (\mu) \, ,
\end{equation}
where $ {\rm P} \, (m \,| \, N, \mu)$ is the likelihood given by Eq.~(\ref{sfxtsims:eq:binomial}) 
and is meant to be a function of $\mu$. The prior is denoted by $p \, (\mu)$.
Apart from a normalization term, the likelihood is the Beta distribution of $\mu$ given $N$ and $m$
\begin{equation}
\label{sfxtsims:eq:beta}
{\rm Beta} \, (\mu \, | \, N, m) = \frac{\Gamma\,(N+2)}{\Gamma\,(m+1) \, \Gamma \, (N-m+1)} \, \mu^{m} \,  (1-\mu)^{N-m}  .
\end{equation}
The convenient choice \citep[][Sect. 2.2.1]{Bishop2006:PRML} for a prior function is a conjugated distribution, 
which is the Beta distribution with parameters $a$ and $b$, 
\begin{equation}
\label{sfxtsims:eq:prior}
p \, (\mu | \, a, b)  = {\rm Beta} \, (\mu \, | \, a, b) = \frac{\Gamma\,(a+b)}{\Gamma\,(a) \, \Gamma\, (b)} \, \mu^{a-1} \,  (1-\mu)^{b-1} .   
\end{equation}
After proper normalization, the posterior in Eq.~(\ref{sfxtsims:eq:posterior}) becomes
\begin{equation}
\label{sfxtsims:eq:posterior2}
{\rm P} (\mu | m, N, a, b)  = \frac{\Gamma\,(N+a+b)}{\Gamma\,(m+a) \, \Gamma\, (N-m+b)} \mu^{m+a-1} \,  (1-\mu)^{N-m+b-1} . 
\end{equation}
From Eq.~(\ref{sfxtsims:eq:posterior2}) the expectation value and variance of $\mu$ are 
\begin{eqnarray}
\label{sfxtsims:eq:expvar1}
 E\{\mu\} & = & \frac {m+a}{N+a+b} = \frac {\mu_{\rm est} + a/N} {1+(a+b)/N} \\
{\rm var}\{\mu\} & = & \frac{(m+a)\,(N-m+b)}{(N+a+b)^2 \, (N+a+b+1)}   \nonumber \\
    & =  & \frac{(\mu_{\rm est} +a/N) \, (1-\mu_{\rm est} +b/N)}{(1+(a+b)/N)^2 \, (N+a+b+1)} . 
\end{eqnarray}
The case of an uninformative prior is easily recovered for $a=b=1$. 
We note that in the asymptotic limit of large values of $N$,
\begin{eqnarray}
\label{sfxtsims:eq:expvar2}
 E\{\mu\} & \simeq & \mu_{\rm est }\\
{\rm var}\{\mu\} & \simeq & \frac{\mu_{\rm est} \,  (1-\mu_{\rm est})}{N } \,\, ,
\end{eqnarray}
in agreement with the asymptotic limit of a normal distribution. 

For a class of sources consisting of a small number of individuals 
($N_{\rm src} \la 50$) the prior $p \, (\mu)$ is unknown, so only an uninformative prior 
can be used in Eq.~(\ref{sfxtsims:eq:posterior}). 
Such is the case of SFXTs  ($N_{\rm src}=10$), which will be detailed in Sect.~\ref{sfxtsims:mcboot_idc}, 
and for which Eq.~(\ref{sfxtsims:eq:posterior2}) can only be used with $a=b=1$. 

To this end, we provide (on-line only) $R$--language, IDL, and C--language programs that, 
given $N$ and $DC$ as calculated according to Eq.~(\ref{sfxtsims:eq:IDC}), 
provides the 68.3\,\%, 95.4\,\%, and 99.7\,\% c.i. for the 
theoretical distribution (Eq.~\ref{sfxtsims:eq:posterior2}). 

On the contrary,  when $N_{\rm src} >50$ the prior can be obtained from 
the observed distribution of the DCs of all sources by fitting it with the Beta function
in Eq.~(\ref{sfxtsims:eq:prior}) with free parameters  $a$ and $b$.  
In this case,  Eq.~(\ref{sfxtsims:eq:prior}) turns out to be particularly useful 
for the newly discovered sources even with relatively few available observations. 
Through the prior, one can further constrain the DC of a new source 
by exploiting the information (the fitted values of $a$ and $b$) 
acquired on the DC distribution derived from the sources of the same class 
previously observed.

        \section{Evaluating duty cycles with Monte Carlo bootstrap simulations} \label{sfxtsims:mcboot_idc}

Once the best available measurement, $DC(N)$, has been obtained from a set of  $N$
independent observations, one needs to assess its associated error. 
The DC determinations obtained by accumulating increasing observing time are not 
independent; therefore,  the dataset cannot be used to directly determine the error on DC. 
Furthermore,  the datasets can be so poor that the hypothesis of normal errors does not apply. 
The standard approach, also validating {\it a posteriori} our derivation in Sect.~\ref{sfxtsims:statistics}, 
is to perform Monte Carlo simulations. 

As a test case, we consider the \sw/XRT monitoring campaigns on the ten SFXTs 
reported in Table~\ref{sfxtsims:tab:final}, discussed in full by \citet[][]{Romano2014:sfxts_paperX} 
who calculate the IDCs according to Eq.~\ref{sfxtsims:eq:IDC}. 
Table~\ref{sfxtsims:tab:final} (Cols.~1--5) reports the main properties of the sample. 
The data were divided in 
{\it i)} yearly campaigns (Y), a casual sampling of the X--ray light curve of an SFXT 
   at a resolution of $P_{\rm samp} \sim 3$--4\,d  over a $\sim 1$--2\,yr baseline  
   (for these, $P_{\rm samp} \ga P_{\rm orb}$); 
and {\it ii)} orbital campaigns (O), that sample the light curve intensively
with $P_{\rm samp} <<P_{\rm orb}$ so that the phase space is uniformly 
observed within one (or a few) $P_{\rm orb}$. Further details can be found in
\citet[][]{Romano2014:sfxts_paperX}. 

In order to determine the expectation value of $IDC$ and its error, we
performed Monte Carlo bootstrap simulations \citep[][]{Efron1979,Efron1994}.  
We created $M=10^4$ simulated data sets, 
drawn from the observed sample of size $N$ 
with a simple sampling (with replacement, or uniform probability). 
We calculated $M$ values of IDCs (simulated sample) according to Eq.~(\ref{sfxtsims:eq:IDC}).  
The simulated sample mean and standard variance 
(Table~\ref{sfxtsims:tab:final}, Col.~9) are 
\begin{eqnarray}
\label{sfxtsims:eq:IDCsim}
\overline{IDC_{\rm sim}} & =& \frac{1}{M}\sum_{k=1}^{M}IDC_{\rm sim}(k) , \\
\label{sfxtsims:eq:IDCs}
 s^2_{\rm sim}&=&\frac{1}{M-1}\sum_{k=1}^{M}(IDC_{\rm sim}(k)-\overline{IDC_{\rm sim}})^2 .   
\end{eqnarray}
In Fig.~\ref{sfxtsims:fig:distro_idc_M10000} we show, superposed on the 
simulated sample distributions (solid red curves),  
the simulated sample mean $\overline{IDC_{\rm sim}}$ (vertical line), 
and the theoretical expectations (dashed green curves) described by 
Eq.~(\ref{sfxtsims:eq:posterior2}).
We find that  $s_{\rm sim}=2.9$--6\,\%  for the yearly campaigns 
and $s_{\rm sim}=8.0$--10.4\,\% for the orbital ones.

The standard c.i., 
defined by the integral of the probability function (i.e.\ the simulated distributions), 
the cumulative probability function, 
\begin{equation}
\label{sfxtsims:eq:CPF}
F(x)=\int_{-\infty}^{x} IDC_{\rm sim}(x^\prime) \, dx^\prime, 
\end{equation}
 can be calculated from 
\begin{eqnarray}
\label{sfxtsims:eq:IDCsim_confidence}
F(x_{1, 1}) = \frac{1-c1}{2};  \,\,\,     F(x_{2,1}) = \frac{1+c1}{2};  \,\,\,   {\rm and} \,  c1 =0.6827, \\
\label{sfxtsims:eq:IDCsim_confidence2}
F(x_{1,2}) = \frac{1-c2}{2};  \,\,\,     F(x_{2,2}) = \frac{1+c2}{2};  \,\,\,   {\rm and} \,  c2  =0.9545,  \\
\label{sfxtsims:eq:IDCsim_confidence3}
F(x_{1,3}) = \frac{1-c3}{2};  \,\,\,      F(x_{2,3}) = \frac{1+c3}{2}; \,\,\,    {\rm and} \,  c3  =0.9973. 
\end{eqnarray}

        \subsection{IDC as a function of sample size} \label{sfxtsims:mcboot_idc_s}

We can now determine the expected IDC value for a given observed sample size  
via additional Monte Carlo bootstrap simulations. 
For each of the sources monitored with yearly campaigns,  
we created $M=10^4$ datasets drawn from the first $S=10, 20, 30, ..., N$ observed points, 
with a simple sampling (with replacement, or uniform probability). 
The simulated sample mean $\overline{IDC_{\rm S}}$ and the standard deviation $s_{\rm S}$ 
were calculated similarly to Eq.~(\ref{sfxtsims:eq:IDCsim})--(\ref{sfxtsims:eq:IDCs}). 

\begin{figure}
\begin{center}
\vspace{-1.5truecm}
\centerline{
\hspace{+0.5truecm}
\includegraphics*[width=10.3cm,height=12cm]{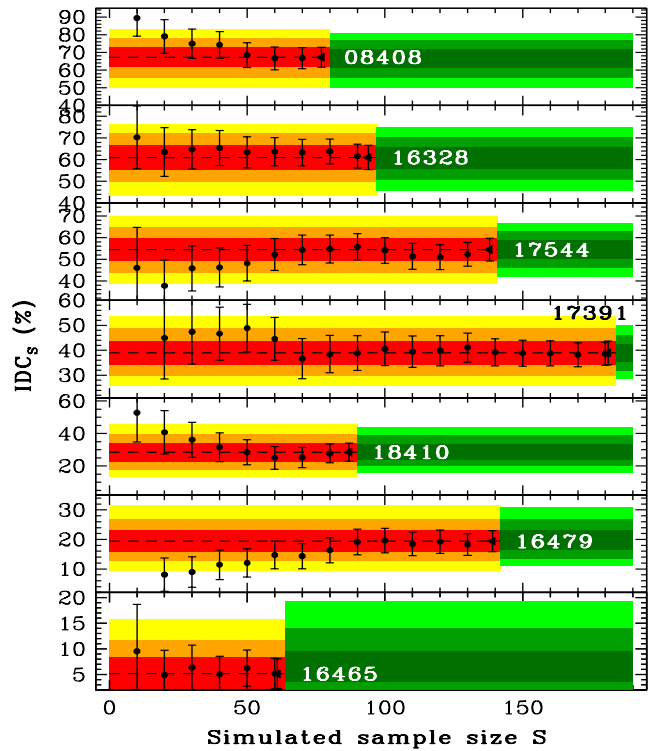}} 
\end{center}
\vspace{-1.truecm}
\caption{
Simulated sample means and their errors $\overline{IDC_{\rm S}}\pm s_{\rm S}$ 
as a function of sample size $S$ (points) for the yearly sample. 
The last point (filled triangle) is the simulation for $N$ points. 
The shaded areas mark the 68.3\,\% (red), 95.4\,\% (orange), and 99.7\,\% (yellow) confidence intervals for $IDC_{\rm sim}$ 
(see Sect.~\ref{sfxtsims:mcboot_idc}) and from Eq.~(\ref{sfxtsims:eq:posterior2}) (dark green, green, light green, respectively; 
see Table~\ref{sfxtsims:tab:final} in the case of a uninformative prior, $a=b=1$). 
}
\label{sfxtsims:fig:idc_s_sig5}
\end{figure}

Figure~\ref{sfxtsims:fig:idc_s_sig5} shows  
$\overline{IDC_{\rm S}}\pm s_{\rm S}$ as a function of the sample size $S.$  
The last point (filled triangle) is the simulation for $N$ points 
for which $\overline{IDC_{\rm N}} = \overline{IDC_{\rm sim}}$ and $s_{\rm N} = s_{\rm sim}$ 
(Eqs.~\ref{sfxtsims:eq:IDCsim} and \ref{sfxtsims:eq:IDCs}). 
The red-orange-yellow bands mark the 68.3\,\%, 95.4\,\%, and 99.7\,\% c.i.\ 
for the simulated distribution as derived from 
Eqs.~(\ref{sfxtsims:eq:IDCsim_confidence})--(\ref{sfxtsims:eq:IDCsim_confidence3}).   
We note the excellent correspondence between the 68.3\,\% c.i.\ (red band) 
and the simulated sample standard deviation $s_{\rm N}$ (the error-bar on the simulation for $N$ points), 
as expected from a normal distribution. 
The green bands (from dark to light green) mark the 68.3\,\%, 95.4\,\%, and 99.7\,\% c.i.\ 
for the theoretical distribution in Eq.~(\ref{sfxtsims:eq:posterior2}) 
also reported in Table~\ref{sfxtsims:tab:final}, Cols.~6--8.

We define $S_a$ as the minimum $S$ value for which $IDC(S)$ is considered acceptable, 
that is the number of observations required in order to satisfy both conditions: 
$$\overline{IDC_{\rm S}} \in  {[\overline{IDC_{\rm sim}}-s_{\rm sim}, \overline{IDC_{\rm sim}}+s_{\rm sim}]}  $$  
$$\overline{IDC_{\rm S}}\pm s_{\rm S} \in  {[\overline{IDC_{\rm sim}}-2\,s_{\rm sim}, \overline{IDC_{\rm sim}}+2\,s_{\rm sim}]} . $$ 
The values of $S_a$ thus determined are reported in Table~\ref{sfxtsims:tab:final}, Col.~10, and they 
range between 40 and 80 observations, depending on the source. 

Similarly, for each of the sources monitored with orbital campaigns,  
we created $M=10^4$ datasets drawn from $S=5, 10, 15, 20, ..., 70$ observed points, 
thus also extrapolating the observed sample 
to determine how many additional observations are required to significantly lower the uncertainty $s_{\rm S}$.  
We find that for about 70 observations $s_{\rm S}=3.6$--5.8\,\%, thus comparable to those 
found for the yearly monitoring campaigns. 

These findings can easily be used for planning future observations.

\section{Conclusions \label{sfxtsims:conclusions}}

 As an alternative and considerably less time-consuming approach than Monte Carlo 
bootstrap simulations, we derived the theoretical Bayesian expectation value 
for a duty cycle and its error based on a finite set of independent observational 
data points. 
We have applied our findings to the specific case of the inactivity duty cycle 
of  SFXTs, as one of the available examples of two-state sources. 
For SFXTs we have compared the theoretical expectations with both the 
observed values and with the IDCs and their errors obtained from Monte Carlo 
simulations, 
as an {\it a posteriori} validation of the Bayesian treatment.

Our treatment, however, is more general than the simple case we considered 
and can be applied to all independent observations of two-state sources, 
such as AGNs, X--ray binaries, etc., suitable for a meaningful DC determination. 
In particular,  the strength of this approach becomes evident 
when considering a well-populated class of sources ($N_{\rm src} \geq 50$) for which,  
the parameters $a$ and $b$ can be obtained by fitting the distribution of the observed DCs 
for all sources in the class with the Beta function in Eq.~(\ref{sfxtsims:eq:prior}), 
thus fully characterizing the prior. 
Then, whenever a new source in the same class is observed for relatively few observations, 
the knowledge of the prior derived from the whole class can be utilized to further constrain 
the DC of this still poorly studied individual source by adopting the $a$ and $b$ of the class.

\begin{acknowledgements}
We thank A.\ Stamerra, P.\ Esposito, V.\ Mangano, and E.\ Bozzo for  helpful discussions. 
CG acknowledges PRIN MIUR project on "Gamma Ray Bursts: From Progenitors to Physics of the Prompt
Emission Process," PI: F.\ Frontera (Prot. 2009 ERC3HT). 
LD thanks Deutsches Zentrum f\"ur Luft und Raumfahrt (Grant FKZ 50 OG 1301).
We also thank the referee for comments that helped improve the paper.
The \sw/XRT data were obtained through target of opportunity observations 
(2007-2012; contracts ASI-INAF I/088/06/0, ASI-INAF I/009/10/0) 
and through contract ASI-INAF I/004/11/0 (2011-2013, PI P.\ Romano).  
\end{acknowledgements}

\end{document}